\documentclass[12pt]{cernrep}
\def \apipi{A_{\pi \pi}}
\def \ocb{\overline{{\cal B}}}
\def \beq{\begin{equation}}
\def \bo{B^0}
\def \cpp{C_{\pi \pi}}
\def \eeq{\end{equation}}
\def \lpp{\lambda_{\pi \pi}}
\def \ob{\overline{B}^0}
\def \rpp{R_{\pi \pi}}
\def \spp{S_{\pi \pi}}
\begin{document}
\title{WEAK COUPLING PHASES}
\author{J. L. Rosner \\
Enrico Fermi Institute and Department of Physics \\
University of Chicago, Chicago, IL, USA}
\maketitle
\begin{abstract}
Recent results obtained from $B$ decays on the phases of weak couplings
described by the Cabibbo-Kobayashi-Maskawa (CKM) matrix are discussed, with
particular emphasis on $\alpha$ and $\gamma = \pi - \beta - \alpha$.
\end{abstract}
\vspace{-2.5in}
\rightline{EFI 02-96}
\rightline{hep-ph/0207197}
\rightline{Presented at CERN CKM Workshop}
\rightline{13--16 February 2002}
\vspace{2in}
\section{INTRODUCTION}

The phases of Cabibbo-Kobayashi-Maskawa (CKM) matrix elements
describing charge-changing weak couplings of quarks are fundamental
quantities.  They are sometimes described in terms of angles $\alpha = \phi_2$,
$\beta = \phi_1$, and $\gamma = \phi_3$ in the unitarity triangle.
Now that BaBar and Belle are converging on a value of
$\sin(2 \beta)$, attention has turned to ways of
learning $\alpha$ and $\gamma = \pi - \beta - \alpha$. This summary
describes some recent work on the subject.

In Sec.\ \ref{sec:bpipi} we discuss $B^0 \to \pi^+ \pi^-$ in the
light of recent measurements at BaBar \cite{Aubert:2002tv} and Belle
\cite{Abe:2002qq} of time-dependent asymmetries.  This work was performed
in part in collaboration with M. Gronau \cite{Gronau:2001cj,Gronau:2002cj,%
Gronau:2002gj} and in part with Z. Luo \cite{Luo:2001ek}.  We then mention
how to learn $\gamma$ from various $B \to K \pi$ decays (Sec.\ \ref{sec:Bkpi},
collaboration with M. Gronau \cite{Gronau:2001cj} and M. Neubert
\cite{Neubert:1998pt,Neubert:1998jq}), $2 \beta + \gamma$ from
$B \to D^{(*)} \pi$ (Sec.\ \ref{sec:BDpi}, collaboration with D. Suprun and
C.-W. Chiang \cite{Suprun:2001ms}), and $\alpha$ and $\gamma$ from
tree-penguin interference in $B \to PP,~PV$ decays, where $P$ is a light
pseudoscalar and $V$ a light vector meson (Sec.\ \ref{sec:int}, collaboration
with C.-W. Chiang \cite{Chiang:2001ir}).  Sec.\ \ref{sec:other} is a short
guide to other recent work, while we summarize in Sec.\ \ref{sec:sum}.

\section{$\alpha$ FROM $B^0 \to \pi^+ \pi^-$ \label{sec:bpipi}}

We regard $\alpha,\gamma$ as uncertain to about $\pi/4$: $126^\circ
\ge \alpha \ge 83^\circ$, $32^\circ \le \gamma \le 75^\circ$
\cite{Gronau:2001cj}, in accord with $122^\circ \ge \alpha \ge 75^\circ$,
$37^\circ \le \gamma \le 80^\circ$ \cite{Hocker:2001jb}.  If $B^0 \to \pi^+
\pi^-$ were dominated by the ``tree'' amplitude $T$ with phase $\gamma
= {\rm Arg}(V_{ub}^* V_{ud})$, the parameter $\lpp \equiv e^{-2 i \beta}
A(\ob \to \pi^+ \pi^-)/A(\bo \to \pi^+ \pi^-)$ would be just $e^{2 i \alpha}$
and the indirect CP-violating asymmetry $\spp = 2 {\rm Im}\lpp/(1 + |\lpp|^2)$
would be $\sin 2 \alpha$.  Here
\beq
\frac{d \Gamma}{d t} \left\{ \begin{array}{c} \bo|_{t=0} \to f \\
\ob|_{t=0} \to f \end{array} \right\} \propto e^{- \Gamma t}[1 \mp \spp
\sin \Delta m t \pm \cpp \cos \Delta m t ]~~~,
\eeq
$\cpp = (1 - |\lpp|^2)/(1 + |\lpp|^2)$, and $\Delta \Gamma \simeq \Delta m/200$
has been neglected.  In the presence of non-zero
$\Delta \Gamma$ one can also measure $\apipi = 2 {\rm Re} \lpp/(1+|\lpp|^2)$.
Since $|\spp|^2 + |\cpp|^2 + |\apipi|^2 = 1$ one has $|\spp|^2 + |\cpp|^2 \le
1$.  However, one also has a penguin amplitude $P$ involving a $\bar b \to \bar
d$ loop transition involving contributions $\sim$ $V_{ud}^* V_{ub}$,
$V_{cd}^* V_{cb}$, and $V_{td}^* V_{tb} = -V_{ud}^* V_{ub} - V_{cd}^* V_{cb}$.
The decay amplitudes are then
\beq
A(\bo \to \pi^+ \pi^-) = -(|T|e^{i \delta_T} e^{i \gamma} + |P|
e^{i \delta_P}),~
A(\ob \to \pi^+ \pi^-) = -(|T|e^{i \delta_T} e^{-i \gamma} + |P|
e^{i \delta_P}),
\eeq
where the strong phase difference $\delta \equiv \delta_P -\delta_T$.
It will be convenient to define $\rpp \equiv \ocb(\bo \to \pi^+ \pi^-)
/\ocb(\bo \to \pi^+ \pi^-)_{\rm tree}$, where $\ocb$ refers to a branching
ratio averaged over $\bo$ and $\ob$.  One may use $\spp$ and $\cpp$ to learn
$\alpha,\delta$, resolving a discrete ambiguity with the help of $\rpp$
\cite{Gronau:2002cj}.  Alternatively, one may directly use $\spp$, $\cpp$,
and $\rpp$ to learn $\alpha$, $\delta$, and $|P/T|$ \cite{Gronau:2002gj,%
Charles:1998qx}.

Explicit expressions for $\rpp$, $\spp$ and $\cpp$ may be found in
\cite{Gronau:2002cj,Gronau:2002gj}.  In \cite{Gronau:2002cj} we estimated
$|P/T| = 0.276 \pm 0.064$ (see also \cite{Beneke:2001ev}), obtaining $|P|$ from
$B^+ \to K^0 \pi^+$ via (broken) flavor SU(3) and $|T|$ from $B \to \pi \ell
\nu$.  Plotting $\cpp$ against $\spp$ for various values of $\alpha$ in the
likely range, one obtains curves parametrized by $\delta$ which establish a
one-to-one correspondence between a pair $(\spp,\cpp)$ and a pair
$(\alpha,\delta)$ as long as $|\delta| \le 90^\circ$.  However, if $|\delta|$
is allowed to exceed about
$90^\circ$ these curves can intersect with one another, giving rise to a
discrete ambiguity corresponding to as much as $30^\circ$ uncertainty in
$\alpha$ when $\cpp = 0$.  In this case, when $\delta = 0$ or $\pi$, one
has $|\lpp|=1$ and $\spp = \sin2(\alpha + \Delta \alpha)$, where
$\tan(\Delta \alpha) = \pm (|P/T| \sin \gamma)/(1 \pm (|P/T| \cos \gamma)$
is typically $\pm 15^\circ$.  One can resolve the ambiguity
either by comparing the predicted $\rpp$ with experiment (see
\cite{Gronau:2002cj} for details) , or by comparing the allowed $(\rho,\eta)$
region with that determined by other observables \cite{Hocker:2001jb}.  An
example is shown in \cite{Gronau:2001cj}.

Once errors on $\rpp$ are reduced to $\pm 0.1$ (they are now about three times
as large \cite{Gronau:2002cj}), a distinction between
$\delta = 0$ and $\delta = \pi$ will be possible when $\spp \simeq 0$,
as appears to be the case for BaBar \cite{Aubert:2002tv}.  For the Belle data
\cite{Abe:2002qq}, which suggest $\spp <0$, the distinction becomes easier;
it becomes harder for $\spp > 0$.  With 100 fb$^{-1}$ at each of BaBar and
Belle, it will be possible to reduce $\Delta |T|^2/|T|^2$ from its present
error of 44\% and $\ocb(\bo \to \pi^+ \pi^-)$ from its present error of
21\% each to about 10\% \cite{Luo:2001ek}, which will go a long way toward
this goal.
In an analysis independent of $|P/T|$ performed since the workshop, the 
somewhat discrepant BaBar and Belle values of $\spp$ and $\cpp$, when averaged,
favor $\alpha$ between about $90^\circ$ and $120^\circ$ (see Fig.\ 1 of
\cite{Gronau:2002gj}).

\section{$\gamma$ from $B \to K \pi$ \label{sec:Bkpi}}

\subsection{$\gamma$ from $\bo \to K^+ \pi^-$ and $B^+ \to K^0 \pi^+$}

We mention some results of \cite{Gronau:2001cj} on information
provided by $B^0 \to K^+ \pi^-$ decays, which involve both a penguin $P'$ and
a tree $T'$ amplitude.  One can use the flavor-averaged branching ratio $\ocb$
and the CP asymmetry in these decays, together with $P'$ information from the
$B^+ \to K^0 \pi^+$ decay rate (assuming it is equal to the charge-conjugate
rate, which must be checked) and $T'$ information from $B \to \pi \ell \nu$
and flavor SU(3), to obtain constraints on $\gamma$.  One considers the
ratio $R \equiv [\ocb(\bo \to K^+ \pi^-)/\ocb(B^+ \to K^0 \pi^+)][\tau_+/
\tau_0]$, where the $B^+/B^0$ lifetime ratio $\tau_+/\tau_0$ is about 1.07. 
Once the error on this quantity is reduced to $\pm 0.05$ from its value of $\pm
0.14$ as of February 2002, which should be possible with 200 fb$^{-1}$ at
each of BaBar and Belle, one should begin to see useful constaints arising
from the value of $R$, especially if errors on the ratio $r \equiv |T'/P'|$
can be reduced with the help of better information on $|T'|$.

\subsection{$\gamma$ from $B^+ \to K^+ \pi^0$ and $B^+ \to K^0 \pi^+$}

One can use the ratio $R_c \equiv 2 \ocb(BB^+ \to K^+ \pi^0)/\ocb(B^+ \to K^0
\pi^+)$ to determine $\gamma$ \cite{Gronau:2001cj,Neubert:1998pt,%
Neubert:1998jq}.  Given the values as of February 2002, $R_c = 1.25 \pm 0.22$,
$A_c \equiv [{\cal B}(B^- \to K^- \pi^0) - {\cal B}(B^+ \to K^+ \pi^0)]/
\ocb(B^+ \to K^0 \pi^+) = -0.13 \pm 0.17$, and $r_c \equiv |T'+C'|/|p'| =
0.230 \pm 0.035$ (here $C'$ is a color-suppressed amplitude, while $p'$ is
a penguin amplitude including an electroweak contribution), and an estimate
\cite{Neubert:1998pt,Neubert:1998jq} of the electroweak penguin contribution,
one finds $\gamma \le 90^\circ$ or $\gamma \ge 140^\circ$ at the $1 \sigma$
level, updating an earlier bound \cite{Gronau:2001cj} $\gamma \ge 50^\circ$.
A useful determination would involve $\Delta R_c = \pm 0.1$, achievable with
150 fb$^{-1}$ each at BaBar and Belle.

\section{$2 \beta + \gamma$ FROM $B \to D^{(*)} \pi$ \label{sec:BDpi}}

The ``right-sign'' (RS) decay $B^0 \to D^{(*)-} \pi^+$, governed by the CKM
factor $V_{cb}^* V_{ud}$, and the ``wrong-sign'' (WS) decay $\ob \to
D^{(*)-} \pi^+$, governed by $V_{cd}^* V_{ub}$, can interfere through
$\bo$--$\ob$ mixing, leading to information on the weak phase $2 \beta +
\gamma$.  One must separate out the dependence on a strong phase $\delta$
between the RS and WS amplitudes, measuring time-dependent observables
\beq
A_\pm(t) = (1+R^2) \pm (1 - R^2) \cos \Delta m t,~~
B_\pm(t) = - 2 R \sin(2 \beta + \gamma \pm \delta) \sin \Delta m t,
\eeq
where $R \equiv |{\rm WS/RS}| = r |V_{cd}^* V_{ub}/V_{cb}^* V_{ud}| \simeq 0.02
r$, with $r$ a parameter of order 1 which needs to be known better.  In
Ref.\ \cite{Suprun:2001ms} we use the fact that $R$ can be measured in
the decay $B^+ \to D^{*+} \pi^0$ to conclude that with 250 million
$B \bar B$ pairs one can obtain an error of less than $\pm 0.05$
on $\sin(2 \beta + \gamma)$, which is expected to be greater than about
0.89 in the standard model.  Thus, such a measurement is not likely to
constrain CKM parameters, but has potential for an interesting
non-standard outcome.  
 
\section{$\alpha$ and $\gamma$ FROM $B \to PP,~PV$ \label{sec:int}}

Some other processes which have a near-term potential for providing
information on tree-penguin interference (and hence on $\alpha$ and $\gamma$)
are the following \cite{Chiang:2001ir}: (1) the CP asymmetries in $B^+ \to
\pi^+ \eta$ and $\pi^+ \eta'$; (2) rates in $B^+ \to \eta' K^+$ and
$\bo \to \eta' K^0$; (3) rates in $B^+ \to \eta K^{*+}$ and $\bo \to \eta
K^{*0}$; and (4) rates in $B^+ \to \omega K^+$ and $\bo \to \omega K^0$.
Other interesting branching ratios include those for $\bo \to \pi^-
K^{*+}$, $\bo \to K^+ \rho^-$, $B^+ \to \pi^+ \rho^0$, $B^+ \to \pi^+
\omega$, and $B^{(+,0)} \to \eta' K^{*(+,0)}$, with a story for each
\cite{Chiang:2001ir}.  In order to see tree-penguin interference at the
predicted level one needs to measure branching ratios at the level of
$\Delta \ocb = (1-2) \times 10^{-6}$.

\section{OTHER WORK \label{sec:other}}

For other recent suggestions on measuring $\alpha$ and $\gamma$, see
the review of \cite{Fleischer:2001zn} and the contributions of
\cite{Gronau:2001ff} on the isospin triangle in $B \to \pi \pi$ ($\alpha$),
\cite{Gronau:1998vg,Atwood:2001ck} on $B^+ \to D K^+$ ($\gamma$),
\cite{Kayser:1999bu} on $\bo \to D K_S$ ($2 \beta + \gamma$),
\cite{Buras:2000gc}
on $\bo \to K \pi$ ($\gamma$), \cite{Fleischer:1999pa} on $\bo \to \pi^+
\pi^-$ and $B_s \to K^+ K_-$ ($\gamma$), and \cite{Gronau:2000md} on
$\bo \to K^+ \pi^-$ and $B_s \to K^- \pi^+$ ($\gamma$).  These contain
references to earlier work.

\section{SUMMARY \label{sec:sum}}

CKM phases will be learned in many ways.  While $\beta$ is well-known now and
will be better-known soon, present errors on $\alpha$ and
$\gamma$ are about $45^\circ$.  To reduce them to $10^\circ$ or less,
several methods will help.  (1) Time-dependent asymmetries in $B^0 \to \pi^+
\pi^-$ already contain useful information.  The next step will come when both
BaBar and Belle accumulate samples of at least 100 fb$^{-1}$.  (2) In
$B^0 \to \pi^+ \pi^-$ an ambiguity between a strong phase $\delta$ near zero
and one near $\pi$ (if the direct asymmetry parameter $C_{\pi \pi}$ is
small) can be resolved experimentally, for example by better measurement of
the $B^0 \to \pi^+ \pi^-$ branching ratio and the $B \to \pi \ell \nu$
spectrum.  (3) Several $B \to K \pi$ modes, when compared, can constrain
$\gamma$ through penguin-tree interference.  This has been recognized, for
example, in \cite{Hocker:2001jb}.  (4) The rates in several $B \to PP,~PV$
modes are sensitive to tree-penguin interference.  One needs to measure
branching ratios with errors less than $2 \times 10^{-6}$ to see such effects
reliably.

\section*{ACKNOWLEDGEMENTS}

I would like to thank C.-W. Chiang, M. Gronau, Z. Luo, M. Neubert, and
D. Suprun for enjoyable collaborations, the workshop organizers for the
opportunity to participate, and the Fermilab Theory Group for hospitality.
This work was supported in part by the United
States Department of Energy under Grant No.\ DE FG02 90ER40560.

\end{document}